\documentclass{aa}
\usepackage{graphicx,natbib}
\bibpunct{(}{)}{;}{a}{}{,}
\usepackage{txfonts}
\newcommand{\flux}{erg~s$^{-1}$~cm$^{-2}$}
\begin{document}
   \title{Unveiling the X-ray/TeV engine in Mkn~421 }

\author{Berrie Giebels \inst{1}
 \and Guillaume Dubus \inst{1,2}
 \and Bruno Kh\'elifi \inst{1}
} \institute{Laboratoire Leprince-Ringuet, UMR 7638 CNRS, Ecole Polytechnique, F-91128 Palaiseau, France \and Institut d'Astrophysique de Paris, UMR 7095 CNRS, Universit\'e Pierre \& Marie Curie, Paris 6, 98 bis bd Arago, F-75014 Paris, France}


\abstract 
	{}
 {Multi-wavelengths observations of TeV blazars provide important clues into particle acceleration in relativistic jets.}
  {Simultaneous observations of Mkn~421 were taken in very high energy $\gamma$-rays ($>200\,\rm GeV$, CAT experiment), X-rays (RXTE) and optical (KVA). Multi-day RXTE observations are also presented, allowing for detailed modelling of the spectral variability.}
	{Short timescale ($\approx 30$~mn) variations in VHE $\gamma$-rays are found, correlated with X-rays but not with the optical. The X-ray spectrum hardens with flux until the photon indices saturate above a threshold flux $\approx 10^{-9}$\flux. The fractional variability decreases from X-rays to optical as a power-law with $F_{\rm var}\propto E^{0.24\pm0.01}$. The full spectral energy distribution is well-fitted by synchrotron self-Compton emission from cooling electrons injected with a Maxwellian distribution of characteristic energy $\gamma_b$. Fluctuations in the injected power with $P_{\rm inj}\propto \gamma_b^4$ explain the observed variability.}
	{The spectral saturation and the power-law dependence of the
	fractional variability are novel results which may extend to other
	TeV blazars. The ability of Maxwellian injections to reproduce the
	observed features suggest second-order Fermi acceleration or magnetic reconnection may play the dominant role in particle acceleration.}
\keywords{Acceleration of particles -- Radiation mechanisms:
  non-thermal -- Galaxies: active  -- BL Lacertae objects: individual:
  Mkn~421 -- Gamma rays: observations -- X-rays: galaxies}

\maketitle

\section{Introduction}

Mkn~421 was the first Active Galactic Nuclei (AGN) to be established
as a TeV emitter \citep{1992Natur.358..477P}. Such very high energy
(VHE) $\gamma$-ray emission is thought to arise from particles
accelerated in a relativistic jet directed along our
line-of-sight. Continuous improvements in Atmospheric Cherenkov
Telescopes (ACT) have resulted in new detections of sources, shedding
new light on the underlying processes, notably in conjunction with
X-ray observatories. Mkn~421, Mkn~501 and 1ES~1959$+$650 are the
brightest VHE extragalactic objects detected so far, with VHE flares
reaching levels many times that of the Crab flux (see
e.g. \citealt{2003ApJ...597..851K}). Bright objects like these can be
studied with ACTs on smaller time scales than fainter objects since
there are more photons.

Variability in Mkn~421 is present at all wavelengths and correlations
during high states have been found from millimeter wavelength
observations up to VHE energies suggesting a single population of
particles may be responsible, if only partly, for radiation over most
of the spectrum. The flux doubling timescales vary widely but are
correlated with energy as the highest energies vary faster.

The spectral energy distribution (SED) has the typical double-humped
blazar feature in $\nu F_\nu$ representation, where the peaks are
separated by 8 to 10 orders of magnitude in energy. The lower
component is attributed to synchrotron radiation from relativistic
electrons/positrons, and peaks at a frequency $\nu_S$ which ranges
from $\approx$ 0.1 keV in a low activity state to a few keV in flaring
cases, being clearly correlated with the intensity
\citep{2004ApJ...601..759T}. The higher component can be
interpreted as inverse Compton scattering of leptons on the
synchrotron photons themselves. An alternative to leptonic models are
the so-called "hadronic models" where the relativistic jet has a
significant component of relativistic protons emitting X-rays and
$\gamma$-rays via synchrotron, photo-production, or pion decay from
proton-proton interactions (see e.g. \citealt{2004NewAR..48..497C} for
a recent review).

Mkn~421 was in a historically high state in the first months of 2001,
triggering a multi-wavelength campaign and spawning a wealth of
observations with unprecedented detection levels in most energy
bands. The campaign included amongst others the ground-based CAT
Cherenkov telescope (200 GeV - 20 TeV), the RXTE X-ray observatory
(2-30 keV) and the 60 cm KVA telescope on La Palma. The RXTE follow-up
campaign was optimised for the Whipple ACT, but some
observations occured during nights at the location of the CAT
observatory. An analysis of the 4 ks segment that happened to exactly
overlap with CAT and its impact on SED modeling, are discussed
here. Multi-day spectral analysis of the X-ray data is also
presented. Observations and analysis are found in \S2, the results are
collected in \S3. A model of the SED is developped in \S4 and the
interpretation of the results is discussed in \S5 before concluding.

\section{Observations and data analysis}
\label{sect:obs}

The data taking, reduction and analysis for CAT and RXTE is
described next. The KVA telescope data shown in this paper have been presented
by \cite{2001ICRC....7.2699S} where the reduction and analysis is
described.

\subsection{CAT\label{cat}}

The CAT (Cherenkov Array at Th\'emis) imaging telescope operated on
the site of the former solar plant Th\'emis in the French Pyr\'en\'ees
from 1996 to 2003. The detector collects Cherenkov light
from the particles in atmospheric air-showers with a 17.8~${\rm m}^2$
dish. The light is recorded by a high definition camera with a
4.8$^{\circ}$ field-of-view, comprised of a central region of 546
phototudes in a hexagonal matrix spaced out by 0.13$^{\circ}$ and of
54 surrounding tubes in two ``guard rings''. A full description of the
detector can be found in \cite{1998NIM...416...278}.

The observations were carried out on the night of 23-24 March 2001
when the source was very active in X-rays. The observations were made
in the so-called \textit{wobble} mode, {\em i.e.} the source position
$\vec{S}$ is shifted by 0.29$^{\circ}$ from the field-of-view center.
Showers pointing to this position are considered as ON-source
data. Showers pointing to the symmetrical position $\vec{S'}$ with
respect to the center are considered OFF data and have been used to
monitor the cosmic-ray background.

The $\gamma$-ray image analysis, \textit{i.e.} the method and the
cuts, is based on the standard CAT procedure
\citep{1998NIM...416...425}. Individual events are compared to abacus
of theoretical average images in order to derive the primary direction
and the energy of events. The extraction of the $\gamma$-ray signal
from the remaining background events is based on a novel statistical
method using a maximum likelihood fit \citep{bruno}. This method gives
an estimation of the number of $\gamma$-rays, ${\rm N}_{\gamma, f}$,
and of the source position, $\vec{S}_f$ given the observational
informations of each shower, the fitted arrival direction, $\vec{D}$,
the fitted shower impact parameter on the ground, $\vec{IP}$ and the
pointing elevation of the telescope, $\theta$. The likelihood
function, $\mathcal{L}$, is defined by the product of the probability
density of each event over all events passing cuts in the field of
view (${\rm N}_{\rm TOT}$):
\begin{displaymath}
\mathcal{L}(\vec{D} | {\rm N}_{\gamma, f},
\vec{S}_f, \vec{IP}, \theta ) =
 \prod_{i=1}^{{\rm N}_{\rm TOT}} p(\vec{D} | {\rm N}_{\gamma, f},
 \vec{S}_f, \vec{IP},\theta )
\end{displaymath}

The probability density $p$ of each event can be separated as a
function of the shower type, i.e. a $\gamma$-ray origin or an
hadronic origin:
\begin{eqnarray*}
p = {\rm N}_{\gamma, f}/{\rm N}_{\rm TOT} \times
p_{\gamma}(\vec{D} |
\vec{S}_f,  \vec{IP}, \theta) + \\
(1-{\rm N}_{\gamma, f}/{\rm N}_{\rm TOT}) \times
p_{h}(\vec{D} | \vec{IP}, \theta)
\end{eqnarray*}
The term $p_{\gamma}$ is the point spread function (PSF) of the CAT
detector. It is determined from Monte-Carlo simulations of
$\gamma$-rays with a power-law distribution in energy and for
different bins in elevation angles. The exact value of the power-law
index is not critical because the PSF extension is dominated by the
lowest energy events. A differential spectrum $d{\rm N}/d{\rm E}
\propto {\rm E}^{-2.7}$ was chosen. The term $p_{h}$ is the
probability to reconstruct a shower arrival direction $\vec{D}$ for a
hadronic shower.

As the true arrival directions of hadronic events (cosmic rays) are
isotropic, $p_{h}$ can be parameterized from data taken in fields of
view without any known $\gamma$-ray source.

The likelihood $\mathcal{L}$ is maximised numerically according to the
parameters ${\rm N}_{\gamma, f}$ and $\vec{S}_f$. For the analysis of
the Mkn~421 data, the source position $\vec{S}_f$ has been fixed to
the AGN position in order to achieve a better $\gamma$-ray
extraction. This statistical method has been tested on simulations and
on data taken on the Crab Nebula. No bias on the reconstructed number
of $\gamma$-ray events is detectable, whatever the signal strength.

The significance of the signal is calculated from ratio of the
maximum likelihood calculated on the data, $\mathcal{L}({\rm
N}_{\gamma, f})$ to the likelihood calculated by assuming that no
signal is present, $\mathcal{L}({\rm N}_{\gamma}=0)$. The exact
formula of the significance estimation is:
\begin{displaymath}
 \sigma = \sqrt{2\times \log{(\mathcal{L}({\rm N}_{\gamma,
 f})/\mathcal{L}({\rm\ N}_{\gamma}=0))} }
\end{displaymath}
It has been checked on simulations that this significance estimation
follows a Gaussian distribution of variance equal to 1 in absence of
any signal.

The characteristics of the cuts described above and this statistical
method for the signal extraction have been computed with a
$\gamma$-ray spectrum of spectral index of $\Gamma$=2.55 ($d{\rm
N}/d{\rm E} \propto {\rm E}^{-\Gamma}$), an integral flux comparable
to the one of the Crab Nebula and at a telescope elevation of
$\theta$=60$^{\circ}$. About 41\% of the $\gamma$-ray events are kept
and the expected significance for 1 hour of observation is
$5.7\,\sigma / \sqrt{\rm 1 hour}$. This method gives better
performances than a method based on a classical subtraction ``ON-OFF''
(e.g. \citealt{1999A&A...350...17D}).

The determination of the spectra and the integral fluxes are made with
the standard CAT method \citep{2001A&A...374...895}. Briefly, the
spectrum extraction is based on forward-folding method in which a
parametrization of the spectrum shape is assumed. And the fluxes are
calculated\ assuming a spectral shape derived by the previous
method. In both determinations, the number of $\gamma$-ray events used
for them is extracted by the maximum likelihood method described
previously. \\

\begin{figure}[htb!]
\begin{center}
\resizebox{\hsize}{!}{\includegraphics{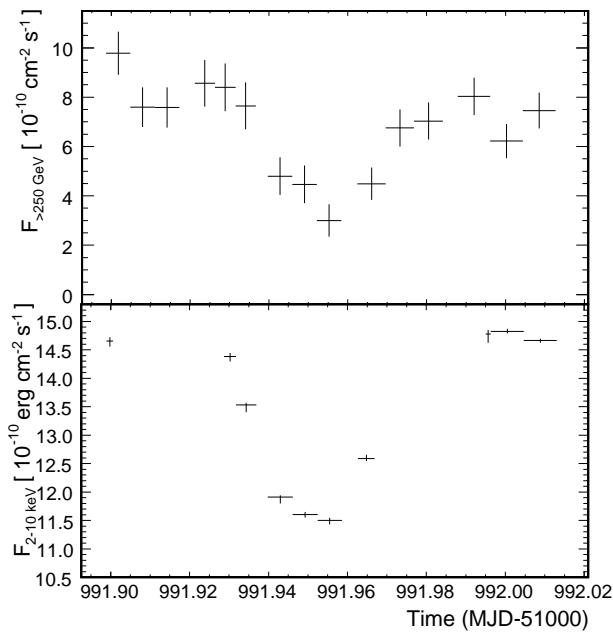}}
\end{center}
\caption{Top: CAT lightcurve on the night of 23 March 2001 expressed
as integrated flux above $200\,\rm{GeV}$. The flux decreases by a
factor $\geq 2$ within 30 min which is the fastest variability seen on
Mkn~421 by CAT. Bottom: RXTE 2-10 keV flux lightcurve for observation
times exactly overlapping or contained within a CAT measurement. The
variability in the VHE band is larger than in the X--rays.}
\label{fig:f1}
\end{figure}

After selecting data based on clear weather conditions and stable
detector operation, a total of 2.3 hours (corrected from dead time) of
observations are kept, all taken within a elevation angle between
70$^{\circ}$ and 90$^{\circ}$. The number of $\gamma$-rays extracted
by the maximum likelihood method are $1362\pm50$ were detected. The
flux was high enough to have significant data points with a 10 minute
sampling. This lightcurve is represented in panel (a) of Figure
\ref{fig:f1}. The integral flux is derived as a function of time
assuming a non-evolving spectrum throughout the night (as it is
justified by the study of the hardness ratio in the section
\ref{sect:var}). A flux decrease of 63\% (significance of 2.3$\sigma$)
within 30 min is followed by a doubling of the flux in 30 min again
(significance of 3.8$\sigma$). These variability timescales had not
been detected by CAT on Mkn~421 before these observations.

The VHE time scale variability of Mkn~421 has been addressed in past
observations. The Whipple observatory's 10 m telescope reported
a $\gamma$-ray rate halving time scale of $\simeq 1\,{\rm h}$ on a
flare in 1999 (\citealt{1999ApJ...526L..81M}), but
this has to be untangled from effective area effects before it can be
assumed that the variability is intrinsic. The fastest doubling time
scale reported by Whipple was 15 min
\citep{1996Natur.383..319G}, also expressed in rate rather than flux,
though it can probably be assumed that effective areas do not double
on this timescale. No reference could be found from Whipple
where the lightcurve is expressed in integrated flux. The HEGRA
collaboration observed the same flaring period but due to poor weather
could not report on this specific episode. A lightcurve derived in
units of integrated flux above 1 TeV shows a halving timescale of
$\leq1\,{\rm h}$ for the night of MJD 51939/40
\citep{2003A&A...410..813A}. The halving timescale reported here is
therefore consistent with other ACT measurements, but it is the
fastest variability seen in an effective area independent way.

\subsection{RXTE}
The X--ray data set used here is publicly available
\footnote{ObsID 60145 which can be obtained from the HEASARC website},
  covering an outburst in Mkn~421 between MJD 51986 and MJD 51994 (the
  whole ObsID has a set of smaller observations that extend until MJD
  52000 but were not included here). The source was detected with a
  rate that required the bright background model.
The high rate allows to segment the data into segments of a few
$100\,{\rm s}$ from which a spectrum and lightcurve expressed in ${\rm
erg}\,{\rm cm}^{-2}\,{\rm s}^{-1}$ can be derived instead of the usual
counting rate. Hence, the correlation between flux and spectral
indices can be studied on the same time scales instead of using rate
vs hardness ratios. This can be particularly relevant if, such as the
case here, the spectrum changes significantly when the brightness
changes.

A small part of the Mkn~421 observation campaign of RXTE is
simultaneous with CAT, corresponding to observations taken between MJD
50991.8 and MJD 50992.1. This segment is analysed such that the RXTE
data segments are made to match the CAT time segments. This means that
the X-ray spectral file from which the flux is derived is at most as
long as the CAT segments from which a flux, hardness ratio or spectrum
was derived, the latter being just one big segment lasting the whole
observation night of 24 March 2001.

The reduction scheme for the X-ray data set is the
following:\begin{itemize}

\item{The STANDARD2 data are extracted in GOODTIME intervals
  constrained to be either a segment at most 400 s long, or at most as
  long as the overlapping length of a VHE observation by CAT, or at
  most as long as the whole CAT data set used to derive the VHE
  spectrum. Due to the variable active PCU configuration during these
  observations, this is done for each PCU independently.}

\item{The PCU dependent PHA files are combined using the {\tt addspec}
tool weighted by the counts information delivered by {\tt
fstatistic} and then the corresponding response matrices were
combined with {\tt addrmf}. The bright background model was used and
only the 3--40 PHA channel range was kept in {\tt XSPEC v. 12.2}, or
approximately 2--$20\,\rm keV$.}

\item{These segments are then fitted to a power-law and a broken
  power-law in XSPEC with PCU configuration-dependent response
  matrices generated by the ftool {\tt pcarsp v. 10.1} and a fixed
  column density of $N_{\rm H} = 1.6\times 10^{20}$ cm$^{-2}$ obtained
  from PIMMS\footnote{See
  http://legacy.gsfc.nasa.gov/Tools/w3pimms.html}. The chisquare
  probability $P(\chi^2)$ for both models is estimated. For the time
  averaged spectrum, a double broken power-law is used.}

\begin{figure}[htb!]
\begin{center}
\resizebox{\hsize}{!}{\includegraphics{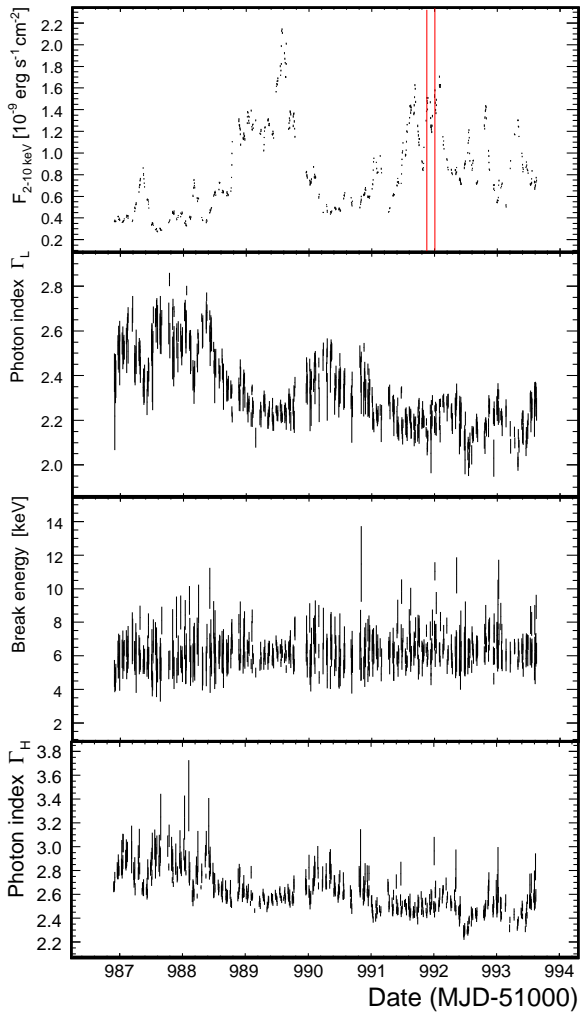}}
\end{center}
\caption{Parameters derived from the broken power-law fits to the
  observation segments, as a function of MJD. The lines in the upper
  panel indicate the segment where simultaneous observations were
  possible with CAT. Whereas there is a clear
  indication of modulation and a flux dependency of both $\Gamma_{\rm
  L}$ and $\Gamma_{\rm H}$, no such indication is visible for the
  break energy.}

\label{fig:fx}
\end{figure}

\item{This yields the flux and the error (defined here as the 1-sigma
  confidence level) on the flux reported in the lightcurve, in units
  of $10^{-11}$ erg s$^{-1}$ cm$^{-2}$ in the 2-10 keV band, as well
  as spectral parameters for each of these intervals, namely the lower
  photon index $\Gamma_{\rm L}$, the break energy $E_{\rm br}$ and the upper
  photon index $\Gamma_{\rm H}$. The broken power-law fit yielded better fit
  results than a single power-law in 97\% of the fits. If the single
  power-law is preferred with a higher $P(\chi^2)$ then both indices
  are set to the single power-law value. Since the data segments do
  not have similar lenghts, the error on the flux can vary but is
  correlated with the integration time (shorter segments have an
  higher associated flux error). A small amount of segments had
  integration times $<200\,{\rm s}$ and were discarded from the
  analysis to keep those with a meaningful statistical error.}

\end{itemize}

The lightcurve derived to match the CAT lightcurve is represented
in panel (b) of Figure \ref{fig:f1}. The total lightcurve for the
analysed proposal 60145 is in panel (a) of Figure \ref{fig:fx} along
with the curve indicating the variation of the two photon indices
$\Gamma_{\rm L}$ and $\Gamma_{\rm H}$ over the same period in panel (b). The
integrated flux lightcurve has also its advantages over a simple
counting rate lightcurve since the rate doesn't properly indicate
how spectral changes affect the flux. Using the photon indices
instead of the hardness ratio gives the absolute spectral
variations. The lightcurve shows tremendous variability, with a
factor of $\simeq$ 10 variation in integrated flux on timescales of
1 d. A similar lightcurve expressed in X--ray counts/second from
this same data set can be found in \citet{2004NewAR..48..419F} along
with the Whipple VHE curve though no spectral X--ray information was
derived in it. A fraction of the same RXTE data set was also
analysed by \citet{2004NewAR..48..395N}.

\section{Results\label{sect:res}}

\subsection{VHE and X--ray Variability\label{sect:var}}

On the night of 2001 March 23, the RXTE/PCA and CAT observations
overlapped exactly for over 2 hours of observation time. During this
time the source underwent a major amplitude variation recorded by both
instruments as seen in Figure \ref{fig:f1}. The amplitude change was
much more dramatic in the VHE band, varying by as much as a factor of
3 whereas the X--ray flux varied by 60\%, implying a difference of a
factor of 5 in flux variability difference.

\begin{figure}[htb!]
\begin{center}
\resizebox{\hsize}{!}{\includegraphics{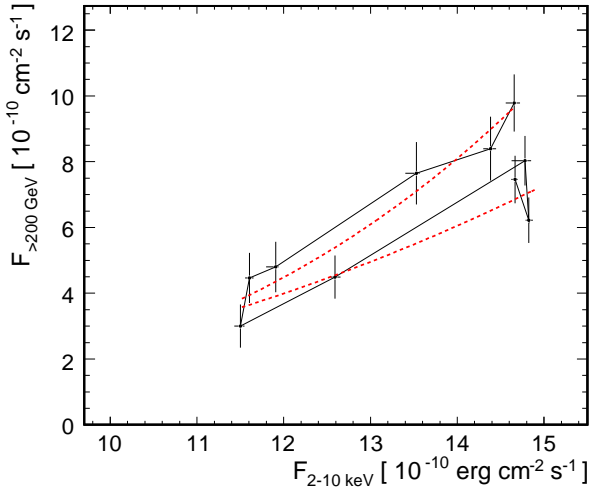}}
\end{center}
\caption{VHE flux versus the X--ray flux for the lightcurve shown in
Figure \ref{fig:f1}. The correlation factor $r=0.86$ indicates that
both wavebands are strongly correlated. There is an indication of a
counter-clockwise pattern in this diagram where the VHE flux is lower
in the rising part than in the decaying phase of the episode seen
here. The decaying and the rising parts of the lightcurve are fitted
to a power-law (dashed lines).}
\label{fig:f2}
\end{figure}

In order to quantitatively look for correlated variability, the VHE
flux is shown versus the X--ray emission in Figure \ref{fig:f2}. The
correlation factor for this data set is $r=0.86$. The
overall correlation plot shows a linear relation between both fluxes
of the form $\Phi ({\rm E>200 GeV})\propto (12.8\pm1.8)\times F_{\rm
2-10keV}$. The amplitude of the variability in the simultaneous light
curves is not large enough show a deviation from a linear
relation, but in order to compare with other results, where
$F_{\rm VHE} \propto F_{\rm X}^\kappa$ relationships are quoted, a
power-law will be used in what follows. 

In order to check any difference between a rising and decaying part of
the emission, the decaying and the rising parts of the lightcurve are
fitted to a power-law, yielding $\kappa = 3.8\pm0.6$ for the decaying
and $\kappa= 2.7 \pm 0.7$ for the rising
parts. Within the error bars no clear difference can be established
between the two. Combining all the points for this transient episode
yields $\kappa = 2.9\pm0.6$.
The Whipple observations from this period were also correlated
with the X--rays. In \citet{2004NewAR..48..395N} the VHE band was
correlated with the ASM rate, where the VHE/X--ray correlation is not
conclusive. This is not surprising since the ASM is less sensitive
than the PCA. \citet{2004NewAR..48..419F} show correlations between
the Whipple $\gamma$-ray rate and RXTE/PCA \textit{counts} with $\kappa=1.3\pm0.3$ for the whole data set
and $\kappa=2.3\pm0.3$ for an individual transient i.e. on a shorter
timescale. The value of $\kappa = 2.9\pm0.6$ derived for the CAT/RXTE
correlation on short timescales is compatible with that result. Note
however that in the case of a changing spectrum, the count rate of the
PCA is not directly proportional to the 2-10 keV flux which is derived
from a spectral fit. \citet{2004ApJ...601..759T} find
$\kappa=1.7\pm0.3$ over a week during the 1998 campaign, using the
peak synchrotron flux determined from ASCA, EUVE and RXTE; and VHE
$\gamma$-ray fluxes in Crab units from HEGRA and Whipple. As
pointed out by \cite{2005A&A...433..479K} on the basis of observations
of both Mkn~421 and Mkn~501, correlating long-term lightcurves may
give results that are significantly different from short-term
transient correlations.

\begin{figure}[htb!]
\begin{center}
\resizebox{\hsize}{!}{\includegraphics{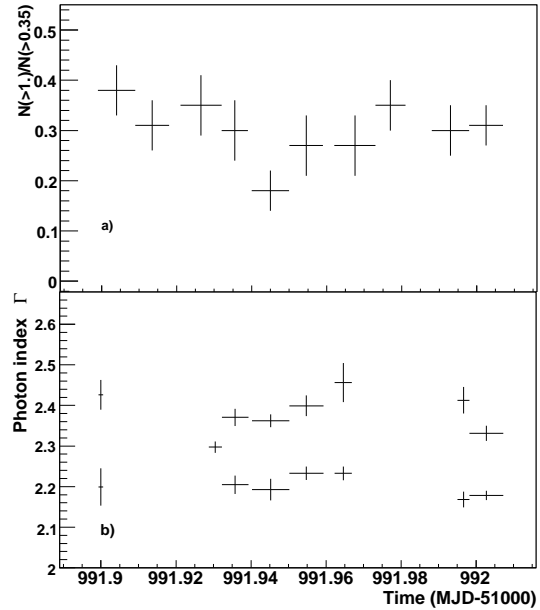}}
\end{center}
\caption{(a): Hardness ratio from the CAT data set, which is
  compatible with a constant. (b): Photon indices $\Gamma_{\rm L}$ and
  $\Gamma_{\rm H}$ on the same time scale as CAT where variability is
  evident for both indices due to their correlation with the flux. No
  correlation is exists between the VHE HR and the X--ray indices.}
\label{fig:f3}
\end{figure}

A hardness ratio HR was derived for the VHE data set, with ${\rm
HR}=\frac{N(>1.)}{N(>0.35)}$ where $N(>1.)$ and $N(>0.35)$ are the
number of $\gamma$-rays with energy above 1 TeV and 350 GeV
respectively. The length of the segments used to derive the ratio
was enlarged to a meaningful statistical significance (panel (a) in
Fig. \ref{fig:f3}). The HR does not appear to be variable for this
episode since it is compatible with a constant function of time
($P(\chi ^2) = 0.13$).

The RXTE data were analysed within these HR time segments, from
which photon indices $\Gamma_{\rm L}$ and $\Gamma_{\rm H}$, the
indices for the power-law component respectively below and above the
break energy $E_{\rm br}$ in the broken power-law, were derived. The
photon index as a function of time is in panel (b) of Figure
\ref{fig:f3}. The variations of $\Gamma_{\rm L}$ and $\Gamma_{\rm
H}$ are not compatible with a constant ($P(\chi ^2) = 0.04$ and
$2.3\times10^{-5}$ respectively) since they are correlated with the
flux.

\subsection{X--ray and optical variability}

\begin{figure}[htb!]
\begin{center}
\resizebox{\hsize}{!}{\includegraphics{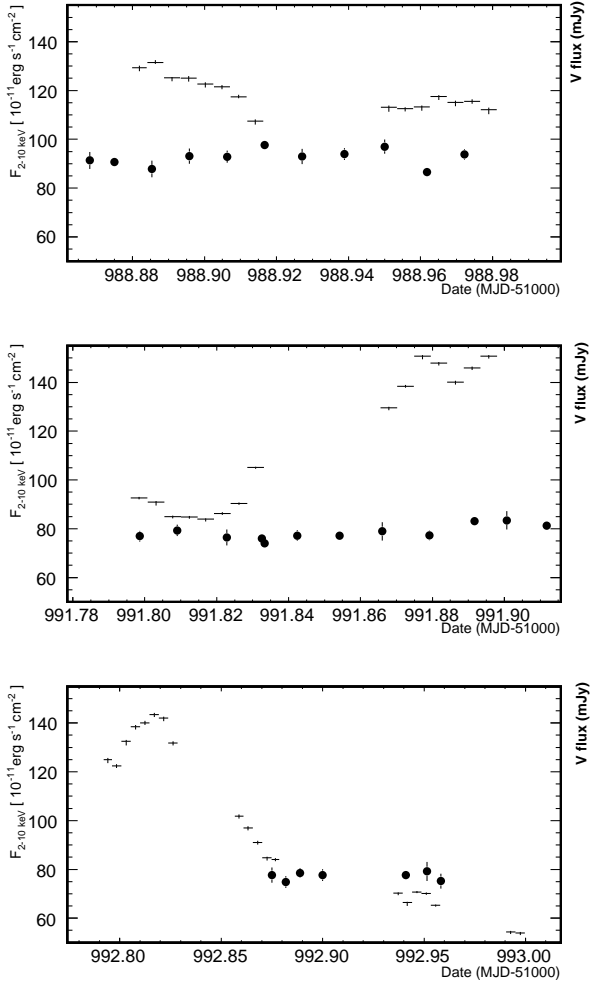}}
\end{center}
\caption{X--ray flux (crosses) and KVA optical flux (points), for 3
  nights where observations overlapped. Note that the Y axis numbering
  is valid for both wavebands. The correlation factor
  $r=0.06$ indicates that both wavebands are not correlated, as is
  apparent in the lightcurves. When the 2-10 keV flux varies by more
  than 50\% no such behaviour is visible in the optical.}

\label{fig:optvar}
\end{figure}

The optical data from the KVA telescope presented in
\citep{2001ICRC....7.2699S} had a few overlapping segments with the
RXTE observations. The lightcurves of Fig. \ref{fig:optvar} show
that there are no counterparts to the rapid and large variations
seen in the X--rays, as has been mentioned in other similar
multi-wavelenght studies. There is hence no evidence for
simultaneous correlation between those wavebands, in agreement with
other multi-wavelength campaigns carried out on this object (e.g.
\citealt{2005ApJ...630..130B}). The absence of correlation between
simultaneous optical/VHE lightcurves is patent in other VHE emitting BL Lacs
such as Mkn~501 \citep{2000ApJ...536..742P}, 1ES~1959+650
\citep{2004ApJ...601..151K} or PKS~2155-304
\citep{2005A&A...442..895A}.

To further investigate the energy-dependent variability, the
fractional root mean square (rms) variability amplitude $F_{\rm var}$ (see
e.g. \citealt{2003MNRAS.345.1271V} for $F_{\rm var}$ and a derivation of
the uncertainty $\sigma_{F_{\rm var}}$ in equation B2) is
estimated for the X--ray and the optical lightcurves. $F_{\rm var}$ is
defined by
\begin{equation}
F_{\rm var} = \sqrt{ \frac {S^2 - \overline{ \sigma^2_{\mathrm err}} }
  {\bar{x}^2}}
\label{eqvar}
\end{equation}
where $s^2$ is the variance of the considered time series,
$\sigma_{\mathrm err}$ is the mean square error of the flux
measurements and $\bar{x}$ the mean of the flux.

\begin{figure}[htb!]
\begin{center}
\resizebox{\hsize}{!}{\includegraphics{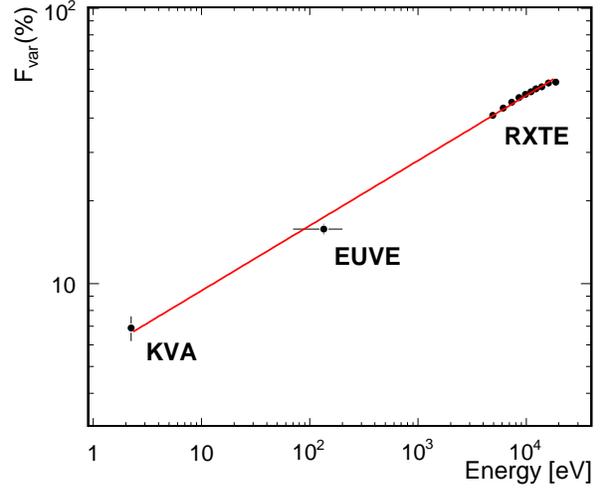}}
\end{center}
\caption{Fractional rms as a function of observed energy, for the RXTE
and KVA data from this campaign. Also a EUVE measurement is shown as
an open circle, from a MWL campaign on Mkn~421
\citep{2000ApJ...542L.105T} in 1998. The data are well described by a
power law with a slope $\propto E^{1/4}$.}
\label{fig:mwlrms}
\end{figure}

Figure \ref{fig:mwlrms} shows $F_{\rm var}$ as a function of energy along
with the optical KVA measurement of $F_{\rm var} =7.55 \pm 0.71\%$, showing
evidence for a power-law behaviour of $F_{\rm var}$ over four decades in
energy with $F_{\rm var}\propto E^{0.24\pm0.01}$. The increase in fractional
variability with energy between 0.1-10~keV was previously noticed by
\citet{2000ApJ...541..153F} using {\em BeppoSAX} data. Their
X-ray fractional rms is also $\propto E^{1/4}$. The historical
EUVE data point shown in Fig.~\ref{fig:mwlrms}, taken from the same
intensive 1998 campaign on Mkn~421, is in very good agreement with the
power-law trend.

\subsection{X--ray spectral patterns}

It is commonly reported in blazar studies that the X--ray flux (or
rate) correlates with the spectral index (or hardness ratio),
indicating that the spectrum becomes harder when the flux increases.
Since the X--ray data analysed here were fitted to a broken power law
the flux dependency of 3 spectral parameters can be checked, namely
the photon indices and the break energy $E_{\rm br}$. Surprisingly the
break energy varies very little with time and also does not correlate
with the flux, a fit to a constant yields a $\chi ^2$ of 11.6 for 9
degrees of freedom, or a 23\% chisquare probability. The distribution
of $E_{\rm br}$ has an average value of $\overline{E_{\rm
br}}=5.9\pm1.1\,{\rm keV}$ independent of the flux.

\begin{figure}[htb!]
\begin{center}
\resizebox{\hsize}{!}{\includegraphics{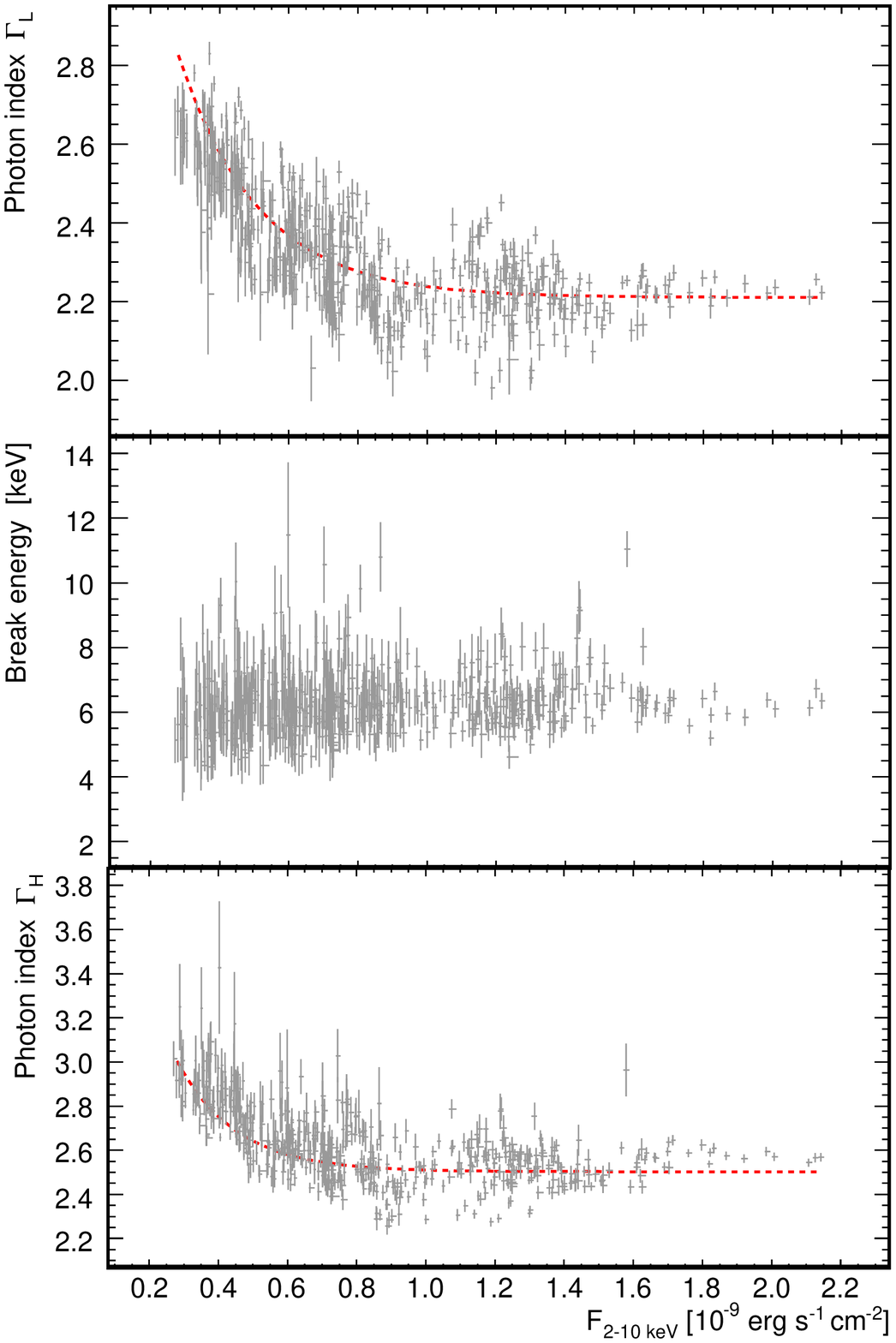}}
\end{center}
\caption{Dependence of the three spectral parameters on the flux, all
derived from fits to a broken power law. The indices are seen to be
levelling at some characteristic flux, whereas the break energy does
not appear correlated with the flux.}
\label{fig:figs}
\end{figure}

The photon indices $\Gamma_{\rm L}$ and $\Gamma_{\rm H}$ however show
some interesting features where the flux dependency is apparent up to
a flux $F_p \approx 10^{-9}\,{\rm erg}\,{\rm cm}^{-2}\,{\rm s}^{-1}$
after which the indices appear to plateau
(Fig. \ref{fig:figs}). Mentions of a possible hardness ratio
saturation were made in \citet{1988MNRAS.232..793G} and
\citet{2004NewAR..48..395N} but the datasets were insufficient to
establish this behaviour.

A similar analysis was performed on Mkn~501 by
\citet{2005ApJ...622..160X} who derived the same broken power-law
spectral parameters for that source and also found that the two photon
indices show a spectral hardening and then level off at a flux of
$F_{\rm p}\approx 5\times10^{-10}\,{\rm erg}\,{\rm cm}^{-2}\,{\rm
s}^{-1}$. The authors used a function consisting of a constant plus an
exponential to describe the trend. This functional form does
also fit quite well the Mkn~421 data shown here, as seen in the dashed
lines in Figure \ref{fig:figs}. 

The two photon indices are correlated ($r=0.99$) and a
linear fit to the data set yields $\Gamma_{\rm H} \propto
(0.96\pm0.02) \times \Gamma_{\rm L}$ showing that the dependency is
very linear, unlike the flux dependency found previously. The ratio
between the two indices remains therefore constant throughout the
whole observation sequence here, and is independent of the measured
intensity. The part of the X-ray spectrum here appears to be
pivoting around the 6 keV break energy, which is probably more an
indication of a spectral curvature than a feature in the spectrum. 

\subsection{Resolved flares}

\begin{figure*}[htb!]
\begin{center}
\includegraphics[angle=0,scale=0.5]{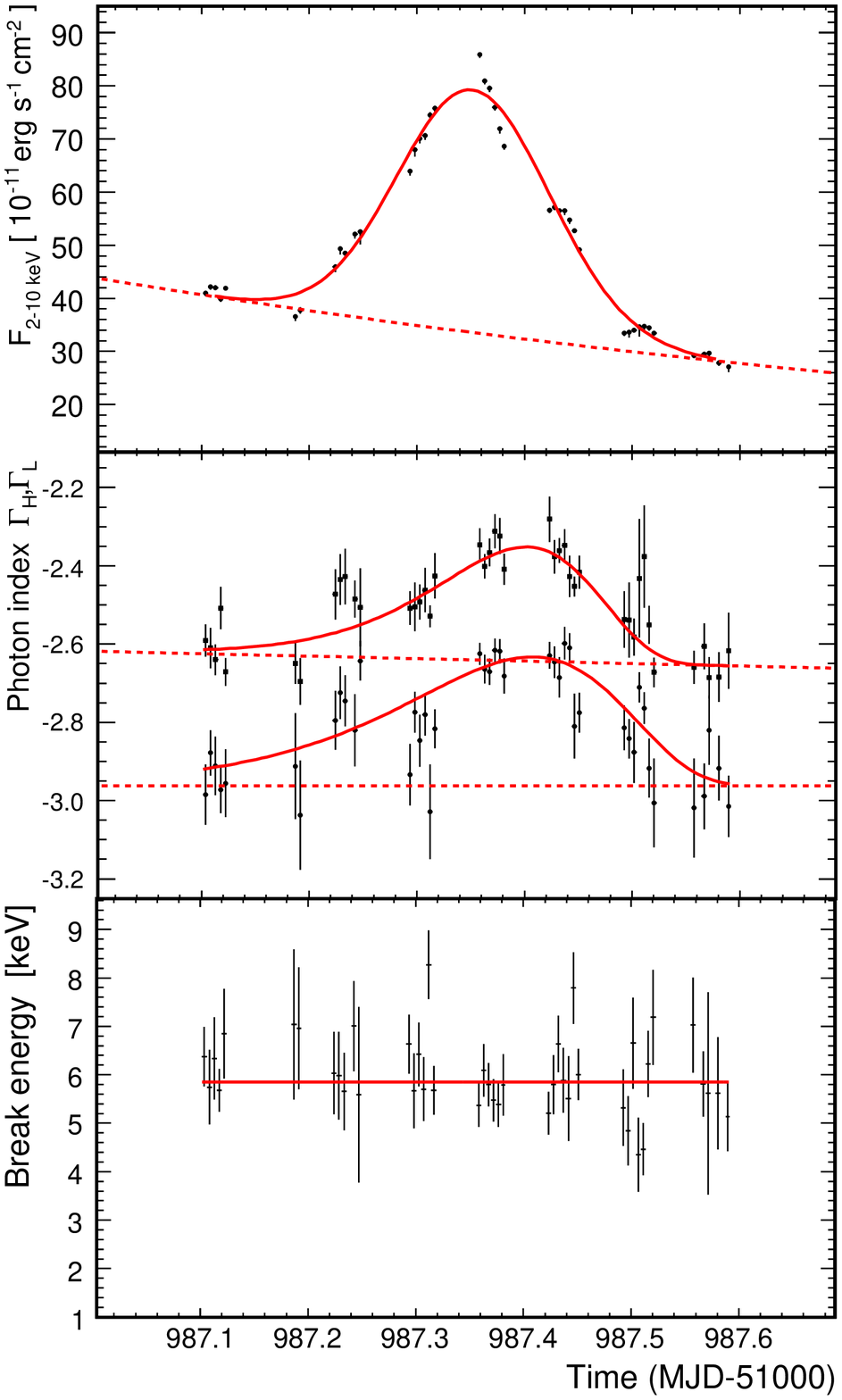}\includegraphics[angle=0,scale=0.5]{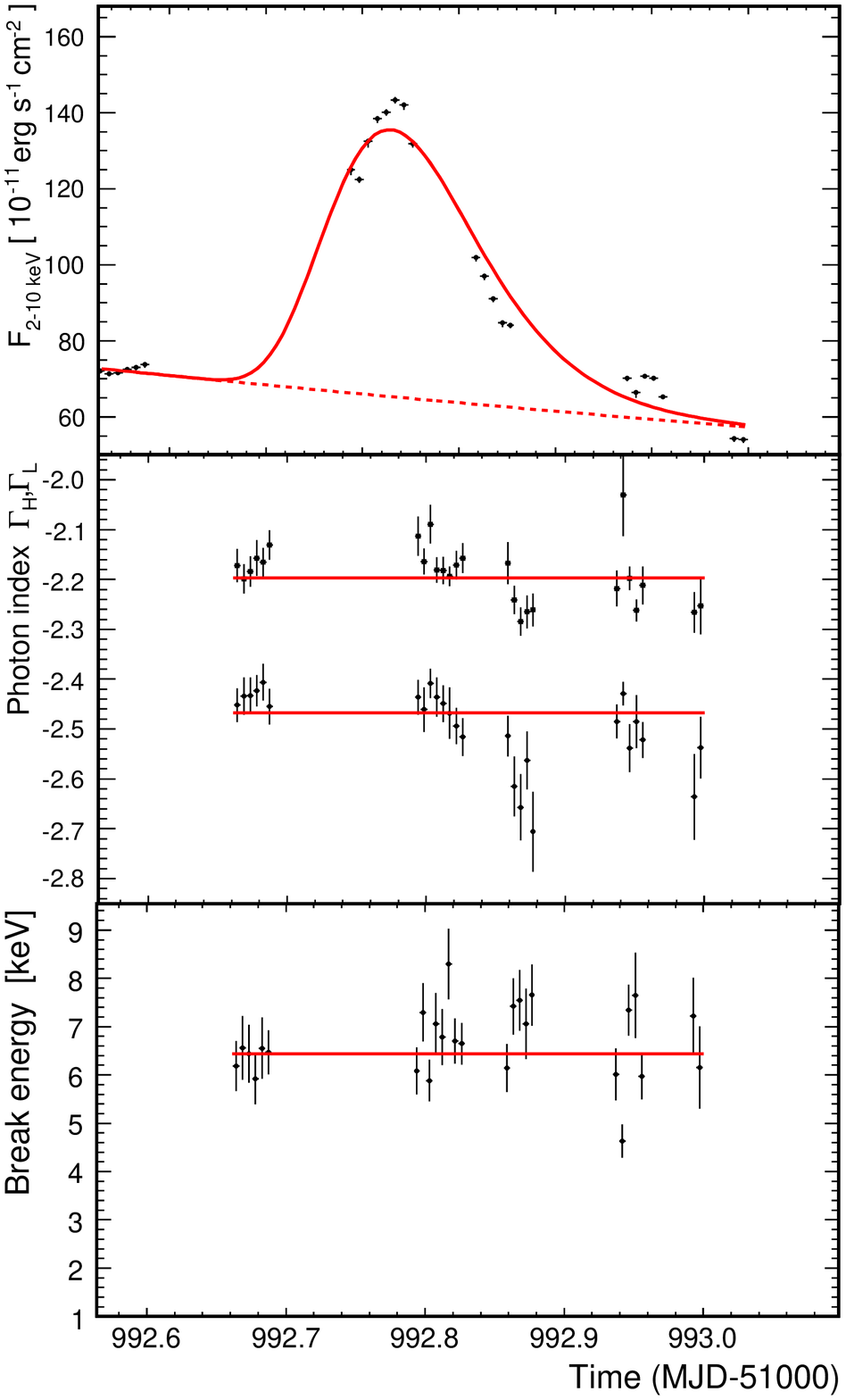}
\end{center}
\caption{Spectral variations for a resolved flare where the flux is
  $\lesssim F_p$ (left panel) and $\gtrsim F_p$ (right panel). In the
  former case, both indices harden from their baseline value, peak
  later than the flux, and soften again with the characteristic
  hysteresis loop. The spectral indices are highest well after the
  flux peaks. In the latter case, no significant spectral variability
  is found when the flux doubles. The lines represent a fit of the
  flares to a lognormal function above a baseline modelled by an
  exponential decay (dashed line). These functions help to locate the
  extrema of the lightcurves but are not used to derive any physical
  parameters.}
\label{fig:f7a}
\end{figure*}

As noted by \cite{2004ApJ...605..662C} the shortest rise or decay
times in Mkn~421 can be as short as 1000~s, but clearly resolved
rising, peaking and decaying structures are not easy to define in the
lightcurve due to the erratic exposure.  There are however two
rising and decaying structures visible in the intervals
MJD 51987.0-51987.7 (flare A) and MJD 51992.6-51993.0 (flare B) with
different amplitudes. Flare A's flux is $\lesssim F_p$ whereas Flare B
is $\gtrsim F_p$. This provides an interesting test for the spectral index
behaviour: both flares start from a different baseline level, double
their flux, and don't exhibit the same spectral evolution: only the
fainter flare shows a clear spectral evolution (Figure
\ref{fig:f7a}). The fact that the spectral index varies more when the 
flux is below the $F_p$ limit, both in the overall lightcurve as well as
in resolved flares, will be discussed further in \S\ref{pvar}.

\subsection{Time averaged spectrum}

The CAT spectrum is derived from data between 300 GeV and 5 TeV, taken
at zenith angles taken in the range $0^\circ
- 20^\circ$. The most probable spectrum for this data is $\frac{{\rm
    d}N}{{\rm d}E} = \Phi_0 \times E^{-\alpha -\beta
\times {\rm log}_{10} E/1{\rm TeV}}$ with
$\Phi_0$=$(13\pm0.85_{stat}\pm 2.76_{syst}) \times 10^{-11} {\rm
cm}^{-1} {\rm s}^{-1} {\rm TeV}^{-1}$, $\alpha$=$
2.81\pm0.09_{stat}\pm0.07_{syst}$ and
$\beta$=$0.79\pm0.27_{stat}\pm0.03_{syst}$. The curvature term is
significant at $2.7\sigma$. The same data set can
be fitted to a power law with an exponential cutoff yielding a similar
likelihood with a cutoff energy of
$E_0=1.4\pm0.5\,{\rm TeV}$. The spectral curvature here is
consistent with HEGRA \citep{2003A&A...410..813A} on
Mkn~421 during these outbursts (though with a different time coverage)
where a spectral cutoff at
$E_0=3.4\pm0.5\,{\rm TeV}$ at the $2.4\sigma$ level was
reported. The Whipple data also show a curved
spectrum \citep{2002ApJ...575L...9K} compatible with an exponential
cutoff at $E_0=4.3\pm0.3\,{\rm TeV}$ as well
as the fact that there is no evidence for variation in the cutoff
energy with flux. The difference in the value for
$E_0$ from these experiments can be due to systematic effects (the
errors quoted here are statistical only, all
experiments give a systematic error on $E_0$ of  $\sim 20\%$) and to
the different observation times.

In order to derive a time-averaged X--ray spectrum consistent with the
VHE spectrum, the RXTE data are reduced to obtain PCU-dependent PHA
files constrained to start at the beginning of the first CAT run, and
to end at the last run. The spectrum is fitted to a double broken
power-law in XSPEC which yields a $\chi^2$ of 52.1
for 50 degrees of freedom, with an excellent chi-square probability of
$P(\chi^2)=0.39$. The spectral parameters are $\Gamma_{\rm
L}=2.20\pm0.06$, $\Gamma_{\rm M}=2.37\pm0.10$, $\Gamma_{\rm
H}=2.64\pm0.04$, $E_{\rm br,1}=6.2\pm0.1\,\rm keV$ and $E_{\rm
br,2}=13\pm1\,\rm keV$ where $\Gamma_{\rm M}$ is the photon index
between the two break energies $E_{\rm br,1}$ and $E_{\rm br,2}$.  
Hence, there is significant curvature in the
time-averaged X--ray spectrum, which is in agreement with earlier
observations of Mkn~421 by telescopes such as ASCA
\citep{2004ApJ...601..759T} and BeppoSAX
\citep{2000ApJ...541..166F,2004A&A...413..489M}. The
2-10 keV flux is $\approx 1.34\times 10^{-9}\,{\rm erg}\,{\rm
cm}^{-2}\,{\rm s}^{-1}$. 

The main facts that will be discussed now can be summarized as follows: \begin{itemize}
\item The X-ray and VHE $\gamma$-ray fluxes are highly correlated
  within a time window of 3h.
\item The optical flux rms, not its mean value, is correlated to the higher
  frequencies through its fractional RMS.
\item The X-ray spectrum hardens with increasing flux only up to some
  characteristic level $F_{\rm p}$ after which it levels off. 
\end{itemize}

\section{Synchrotron self-Compton modelling of Mkn~421 \label{sect:dis}}

VHE emission from BL Lac objects is typically interpreted as due to
radiation from particles accelerated to high energies in a
relativistic jet. The nature of the particles, their location(s), the
acceleration mechanism(s), the radiative processes involved, etc, are
all subject to debate. The simplest view is that of synchrotron
self-Compton emission from electrons (positrons) located in a single,
homogeneous zone. This view is adopted in the present work and
confronted to the results accumulated in the previous section.

\subsection{Constraints from the variability\label{subsec:ssm}}

Several simple constraints on the physical conditions in the emission
zone may be obtained from the observations ({\em e.g.}
\citealt{1996A&AS..120C.503G}). The observed variability timescale sets
an upper limit to the emitting zone. The VHE doubling timescale of 30
minutes (Fig.~\ref{fig:f1}) implies
\begin{equation}
R \delta^{-1} \leq 5.4\times10^{13}\,{\rm cm} \label{eq:eq1}
\end{equation}
where $R$ is the characteristic size of the region (assumed spherical
in what follows) and $\delta$ is the boost factor from the bulk
relativistic motion of the jet. The correlated X--ray and VHE
emissions indicate that the emitting particles are likely to be
co-spatial, and hence that the emitting volumes are similar. This is
consistent with the hypothesis of a homogeneous emission zone.

The detection of $\gamma$-rays in the $0.3-10\,\rm TeV$ range implies
that the opacity to pair production $\tau_{\gamma\gamma}$ within the
emission region cannot significantly exceed 1, independently of the
emission mechanisms. Following \citet{1998ApJ...509..608T} the opacity
$\tau_{\gamma\gamma}$ is
\begin{equation}
\tau_{\gamma\gamma} \approx \frac{\sigma_{\rm T}}{5} \frac{1}{hc}
  \frac{d_L^2}{R} \frac{1}{\delta^3(1+z)} F\left( E_t
  \frac{\delta^2}{(1+z)^2}\right).
\label{eq:eq5}
\end{equation}
In the above, $\tau_{\gamma\gamma}$ is the opacity seen by VHE photons
with an observed energy $E_\gamma$. The VHE photons see a flux density
$F(E_t)$ of photons above the threshold for pair production
$E_t=(m_ec^2)^2/E_\gamma$, which occurs in the UV to X--ray range
(observer frame). If the emission is co-spatial, the measured X--ray
data can be used to estimate the flux density that VHE $\gamma$-rays
may interact with. Imposing that $\tau_{\gamma\gamma}<1$ for photons
with observed energy $E_\gamma=10\,\rm TeV$ and using Eq. \ref{eq:eq1}
to constrain $R$ yields
\begin{equation}
\delta\geq16 \label{eq:eq7}.
\end{equation}

The mean energy $E$ of synchrotron radiation from an electron with
Lorentz factor $\gamma$ is given by $E \approx
\frac{\delta}{1+z}\frac{21\hbar\gamma_{\rm e}^2qB}{15\sqrt{3}m_e}$
where $B$ is the magnetic field intensity in the emission zone.  If
the observed X-rays are synchrotron radiation from electrons then the
maximum observed energy $E_{\rm max}=20\,\rm keV$, corresponding to the 
highest
energy bin in the RXTE spectrum, yields
\begin{equation}
\gamma_{\rm max} \ga 1.4\times10^{6}\, B^{-1/2}\delta^{-1/2} 
\label{eq:eq8}
\end{equation}

In the SSC scenario, the same electrons should have enough energy to
Comptonize ambient synchrotron photons up to the VHE regime, requiring
\begin{equation}
\frac{\delta}{1+z}\gamma_{\rm max} m_ec^2 \approx 10\, {\rm TeV.}
\label{eq:eq9}
\end{equation}
Combining Eq. \ref{eq:eq8} and Eq. \ref{eq:eq9} yields a constrain on the 
magnetic field
\begin{equation}
B\delta^{-1} \leq 5.2\times10^{-3}\, {\rm G} \label{eq:eq11}
\end{equation}
which, for $\delta=16$, would imply $B\leq0.08$~G and $\gamma_{\rm
max}\ga 3.5\times 10^5$. This is of the order of the
Lorentz factor derived in a similar way for Mkn~421 by
\citet{1996ApJ...470L..89T} who derived $\gamma_{\rm max}>5\times10^5$
but lower than $\gamma_{\rm max}\approx 10^7$ found in a similar way
in 1ES1959$+$650 \citep{2002ApJ...571..763G}. Accommodating a higher
magnetic field would require large relativistic speeds (boost factor
$\delta$).

The relative strength of the inverse Compton (IC) contribution
relative to the synchrotron `bump' depends on the particle and
magnetic field densities. Higher densities increase the IC emission
while a higher $B$ increases the synchrotron contribution. In the
Thomson regime, the relative luminosities are given by
\begin{equation}
\frac{L_C}{L_S}\approx \frac{(\nu F_{\nu})_S (d_L/R)^2 / 
(c\delta^{4})}{B^2/8\pi}\ga 810 ~\delta_{16}^{-8} \label{eq:eq12}
\end{equation}
where $B$ and $R$ are replaced by their maximum allowed value
(Eqs.~\ref{eq:eq1} and \ref{eq:eq11}), $(\nu F_\nu)_S\approx
2 \times 10^{-9}$ \flux\ is the observed synchrotron spectral flux,
$d_L\approx 4~10^{26}$~cm is the luminosity distance, and the
numerical application takes $\delta=16$ (Eq.~\ref{eq:eq7}). The small
radius (high particle density) and low magnetic field implied by the
constraints lead to a towering IC component, while both components are
observed at roughly the same spectral luminosities. Therefore, this
leads to values of $\delta$ closer to $\approx 40$ to match the
relative luminosities.

In Mkn 421, inverse Compton emission at VHE energies occurs in the
Klein-Nishina (KN) regime, where the electron gives away nearly all
its energy to the photon. This regime is reached when the boosting
electon has a Lorentz factor $\gamma$ such that $\gamma e'_{S} \geq 1$
where $e'_{S}$ is the energy of the target photon (in units of $m_e
c^2$) in the jet flow frame. With synchrotron peaking at $\approx
500$~eV then $e'_{S}\approx 10^{-3} \delta^{-1}$ and IC on electrons
with $\gamma \geq 10^3 \delta$ occurs in the KN regime. The IC
cross-section then decreases dramatically so that the luminosity in
Eq.~\ref{eq:eq12} is overestimated. Nevertheless, the conclusion that
a high value of $\delta$ is needed to fit the IC component should
still hold: in the KN regime, the electrons will lead to a jet-frame
peak IC luminosity $e'_{IC}\approx 1/e'_{S}$
\citep{2003ApJ...594L..27G}, which translates into $\delta\geq 30$
since $e'_{IC}\approx 10^6 \delta^{-1}$ (corresponding to about
0.5~TeV in the observer frame).

Hence, the VHE variability timescale and the SED imply large boosts
$\delta\ga 30$, $R\sim 10^{15}$~cm and $B\sim 0.1$~G. 

\subsection{The standard SSC model \label{static}}

The validity of the above limit on $\delta$ is investigated using a
numerical code calculating the homogeneous SSC emission from a given
distribution of electrons in spherical geometry
\citep{1974ApJ...188..353J,1979A&A....76..306G,1985ApJ...298..128B}. 
Radiative
transfer is treated as in \citet{1999ApJ...514..138K}. The Bessel
functions, needed for the synchrotron emission and absorption
coefficients \citep{rybicki}, are computed using the Chebyshev
expansion of \citet{macleod}. The Compton emissivity is computed using
the \citet{1968PhRv..167.1159J} approximating kernel, which takes into
account the reduction of the cross-section in KN regime and assumes an
isotropic photon field. Absorption of VHE $\gamma$-rays due to pair
production with the infrared extragalactic background light (EBL)
during their propagation is taken into account. The pair production
cross-section is given in \citet{1967PhRv..155.1404G}. The
distribution and normalisation of the EBL are set to the upper limits
found by the H.E.S.S. collaboration using observations of more distant
TeV blazars \citep{hessnature}.

The shape of the electron distribution is assumed to be a power-law
with an exponential cutoff ${\rm d}N_\gamma \propto \gamma^{-s}
\exp(-\gamma/\gamma_b) {\rm d}\gamma$ from $\gamma_{\rm min}=1$ to
$\infty$. The observed SED (Fig. \ref{fig:sed2}) is power-law like from radio to optical
with a spectral index $\alpha\approx 0.35$ ($S_\nu \propto
\nu^{-\alpha}$). If the emission is synchrotron then the underlying
electron power-law has an index $s\approx 1.7$. The decrease in
spectral luminosity at X-rays implies a peak (in $\nu F_\nu$)
synchrotron frequency somewhere in the extreme UV range
\citep{2004ApJ...601..759T}. The corresponding $\gamma_b$ is
univocally determined by adjusting to the data once $B$ and $\delta$
are set. It is typically of order $10^5$ as $\gamma_b\la \gamma_{\rm
max}\approx 5~10^5 B^{-1/2}_{0.2} \delta_{40}^{-1/2}$
(Eq.~\ref{eq:eq8} with $B=0.2$~G and $\delta=40$). Besides the shape
of the particle distribution, the model parameters are the magnetic
field $B$, the radius $R$, the boost $\delta$ and the particle density
$n$. The first two are set to their maximal value allowed by
constraints, depending on $\delta$ (Eqs.~\ref{eq:eq1} and
\ref{eq:eq11}). This ensures IC emission is maximised. The only free
parameters left are $\delta$ and $n$. Fits are attempted starting with
$\delta=16$ and going upwards. Adequate fitting of the SED does
require values above $\delta\approx 40$ as argued in \S\ref{subsec:ssm}. 

The information gained by such modelling is limited. The particle
distribution adjusted to fit the observations has $s\approx 1.7$ when
values of 2--2.2 are the canonical values preferred theory-wise
\citep{2001MNRAS.328..393A}.  Using $s\approx2$ would require
$\gamma_{\rm min}\sim 10^3$ ({\em i.e.} a limited range of non-thermal
particles) in order not to overestimate the IR spectrum and enhance
the HE $\gamma$-ray spectrum above the EGRET measurement. A high
$\gamma_{\rm min}$ is not improbable as $\gamma_{\rm min}\sim
\Gamma^2$ is expected for relativistic shock acceleration. Yet, one
parameter is frozen (the index $s$) only to thaw another one
($\gamma_{\rm min}$) at the price of reproducing {\em less} of the
SED: the exercise is not particularly enlightning.

The largest limitation is that such simple modelling is, by
construction, unable to address the wealth of variability information
gathered in the previous sections. Injection and cooling of
particles has to be explicitely taken into account and compared
to the spectral variability.

\subsection{Evolved SSC model: steady-state \label{sstate}}

Particles are now continuously injected in a homogeneous sphere,
losing energy to radiation until they escape after a time
$t_{\rm esc}$ \citep{1997A&A...320...19M,1999MNRAS.306..551C}. The
evolution of the particle distribution $N(\gamma,t)$ is given by the
diffusion equation
\begin{equation}
\frac{\partial N(\gamma,t)}{\partial t}=\frac{\partial}{\partial 
\gamma}\left[ (\dot{\gamma}_{\rm s}+\dot{\gamma}_{\rm ic}) N(\gamma,t)  
\right]+\frac{N(\gamma,t)}{t_{\rm esc}}+Q(\gamma)
\end{equation}
where $\dot{\gamma}$ represents the radiative loss term to synchrotron
and IC radiation, $N/t_{\rm esc}$ is the energy-independent escape
term and $Q(\gamma)$ is the injected particle distribution. The
radiative losses $\dot{\gamma}$ are computed from the synchrotron and
IC power emitted by each electron, taking the Klein-Nishina drop in
cross-section into account. The equation is solved using the
\citet{changcooper} algorithm \citep{1996ApJS..103..255P}. In all that
follows the parameters are set to $\delta=40$, $B=0.2$~G and
$R=2 \times 10^{15}$~cm.

In the above, cooling will be unimportant if the particles escape
before emitting a significant fraction of their energy. Taking into
account only synchrotron radiation, the critical energy above which
cooling is important is
\begin{equation}
\gamma_{\rm cool}=3\times 10^5 ~B_{0.2}^{-2} (ct_{\rm esc}/R)^{-1}.
\label{eq:cool}
\end{equation}
Low-energy particles radiate away most of their energy if they are
given the time. It can also be inferred from this equation that light
travel time delays must be taken into account when investigating the
emission of electrons with $\gamma\ga 3\times 10^5$, as these should
cool on a time $\la R/c$ according to Eq.~\ref{eq:cool}
above.

If the escape time is of order $R/c$ then the steady-state
distribution is essentially the injected one because none of the
particles cool significantly before escaping ($\gamma_{\rm cool}\sim
\gamma_{\rm max}$). The situation is identical to
\S\ref{static}. However, the gyroradius $R_L\approx 10^{10}\, \gamma_6
B_{0.2}^{-1}$~cm of TeV electrons is very small compared to $R$
(Eq.~\ref{eq:eq1}), so it is more likely to have the electrons diffuse
for some time in the emission region rather than see them escape on a
direct flight. If $t_{\rm esc}\gg R/c$ the impact of cooling on the
SED will be important.

 Classically, with a power-law injection of particles $Q\propto
\gamma^{-s}$ and $\dot{\gamma}\propto \gamma^2$ (synchrotron and
Thomson IC), the steady-state ($t\rightarrow \infty$) distribution is
a broken power-law with an index $s+1$ above $\gamma_{\rm cool}$. If,
as diffusive shock acceleration suggests, $s=2$ up to some
$\gamma_{\rm max}$, the cooled spectrum has $s=3$ and a flat SED from
$h\nu_{\rm cool}\approx 13 ~\delta_{40} B_{0.2}^{-3} (ct_{\rm
esc}/R)^{-2}$~keV up to $h\nu_{\rm max}$ ($>$20~keV,
\S\ref{subsec:ssm}). A flat SED hardly fits the optical to X-ray
range: \citet{2004ApJ...601..759T} find the emission is not flat but
peaks in the UV. In addition, the emission below $\nu_{\rm cool}$ would
still be a power-law with $\alpha=0.5$, inconsistent with
observations.

The core problem in reproducing the SED with cooling power-laws is the
large range of injected energies. There are too many low-energy
electrons producing too much flux at optical frequencies and
below. One remedy is to arbitrarily harden the injection to $s=1.7$ as
done in \S\ref{static}, fine-tuning $\gamma_b$ so that the optical to
X-ray transition is reproduced. As previously argued, this is not very
satisfying. Another is to keep $s=2$ but increase $\gamma_{\rm min}$
so that high-energy particles carry more of the total injected energy,
cutting off the low frequency emission. In the strong cooling regime
where $\gamma_{\rm min}\geq \gamma_{\rm cool}$ there is no uncooled
component in the distribution.

Injected particles over a limited energy range was found to work very
well. In the limit of a mono-energetic injection, the steady-state
particle distribution yields a power-law of index 1.7 for the chosen
parameters, in excellent agreement with the observed SED. For
synchrotron and Thomson cooling, the steady-state is well-known to be
a power-law with an index 2. Here, inverse Compton cooling occurs in
the less efficient KN regime, leading to a hardened equilibrium
solution. That the mono-energetic injection naturally yields the right
index without fine-tuning makes this solution extremely attractive.

In the following, relativistic Maxwellian injections \citep{2004ApJ...616..136S,2006A&A...453...47K}
\begin{equation}
Q(\gamma)=K \gamma^2 \exp(-2 \gamma/\gamma_b)
\label{max}
\end{equation}
were adopted instead of strictly mono-energetic injections or
power-laws over very limited ranges\footnote{Note that this is not an
$s=2$ accelerated power-law with exponential cutoff, $\gamma^{-2}
\exp(-\gamma/\gamma_b)$, as previously discussed.}. The results are
identical since the Maxwellian is strongly peaked at its
characteristic energy $\gamma_b$. Maxwellians may arise from second
order Fermi scattering on plasma waves or from acceleration during
magnetic field reconnection (see \S5).

\begin{figure}[htb!]
\begin{center}
\resizebox{\hsize}{!}{\includegraphics{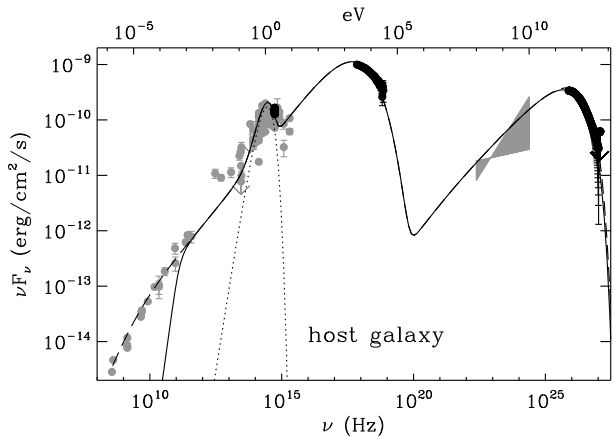}}
\end{center}
\caption{Simultaneous SED of Mkn~421 with archival EGRET (grey
butterfly) and NED (grey points) data. The simultaneous VHE (CAT),
X-ray (RXTE) and optical (KVA) data (\S2) are highlighted in
black. Most of the variability in the synchroton component occurs at
X-ray energies (see Fig.~\ref{fig:mwlrms}) so that archival data below
1~eV, even if non-simultaneous, still provides valid constraints. The
fit (solid line) is a steady-state homogeneous model assuming
continuous injection of a Maxwellian distribution of electrons with
$\gamma$=$5.3\times 10^4$ at a rate $n_{\rm inj}=9.7\times 10^{-7}$
particles~cm$^{-3}$~s$^{-1}$ and cooling via synchrotron and
self-Compton emission. The contribution of the host galaxy to the
optical-IR (dotted line) is modelled using a blackbody spectrum as in
\citet{2003A&A...410..101K}. Values for the other parameters are
$\delta=40$, $B=0.2\,{\rm G}$, $R=2\times 10^{15}\,{\rm cm}$ and $t_{\rm esc}=5000\,R/c$. The full line takes into account pair production on the EBL at
VHE energies and synchrotron self-absorption at radio frequencies. The
dashed line shows the unabsorbed model. Cooling in KN regime of the
injected electrons leads to a steady-state distribution with the right
slope to reproduce the radio to optical slope.}
\label{fig:sed2}
\end{figure}

Figure~\ref{fig:sed2} shows the result of a fit to the simultaneous
optical (KVA), X-ray (RXTE) and VHE (CAT) data, along with NED
archival data and the time-averaged spectral measurement from the
third EGRET catalog \citep{1999ApJS..123...79H}. The contribution of
the host galaxy at optical/IR was modelled following
\citet{2003A&A...410..101K} as a blackbody of temperature $T$=3500~K,
normalised to 7$
\times 10^{-25}$ \flux\ at its peak. The injected
Maxwellian distribution has $\gamma_b$=$1.6\times 10^5$ with $K=3.3
\times 10^{-21}$ in Eq.~\ref{max}. This is equivalent to an injection
rate of $n_{\rm inj}=9.7\times 10^{-7}$ particles~cm$^{-3}$~s$^{-1}$ or a
total injected power $P_{\rm inj}=4\times 10^{39}$~erg~s$^{-1}$.

In the strong cooling regime, the escape time sets the minimum energy
up to which particles cool, given by $\gamma_{\rm cool}$ (taking both
IC and synchrotron into account). With $t_{\rm esc}=50~R/c$ the
synchrotron emission cuts off below optical and is unchanged
above. The VHE emission is also unchanged but the GeV flux is lower as
there are fewer low-energy seed photons. The adopted escape time is
$t_{\rm esc}=5000~R/c$ so that particles cool down to $\gamma\sim 10$
before escaping, emitting low-frequency radio-waves.

This radio emission is self-absorbed in a one-zone model. If the radio
emission zone is more extended (keeping the same $B$), the density of
particles may be low enough to forego self-absorption. In general, the
interpretation of the radio emission from blazars in SSC models
requires the addition of an additional large-scale component
representing the VLBI jet \citep{2006astro.ph..7258C}.

The electron density in steady-state is $n_{\rm inj}t_{\rm esc}\approx
325$ particles~cm$^{-3}$. The energy density in particles is about 5$\times
10^{-2}$\, erg~cm$^{-3}$ compared to a magnetic field energy density
of 1.6$\times 10^{-3}$\, erg~cm$^{-3}$. A sub-equipartition magnetic field
was also found by \citet{1999ApJ...526L..81M} and appears generic of
homogeneous models. The total luminosity is $\approx 1.5\times 10^{46}$\,
erg~s$^{-1}$.

In conclusion, a steady-state homogeneous model with a Maxwellian-like
injection of particles provides a simple description of the overall
SED, removing the need to set-up a power-law with an arbitrary hard
index. In the following, this picture is confronted with the
constraints on variability derived in \S3.

\subsection{Evolved SSC model: variability \label{pvar}}

\begin{figure}[htb!]
\begin{center}
\resizebox{\hsize}{!}{\includegraphics{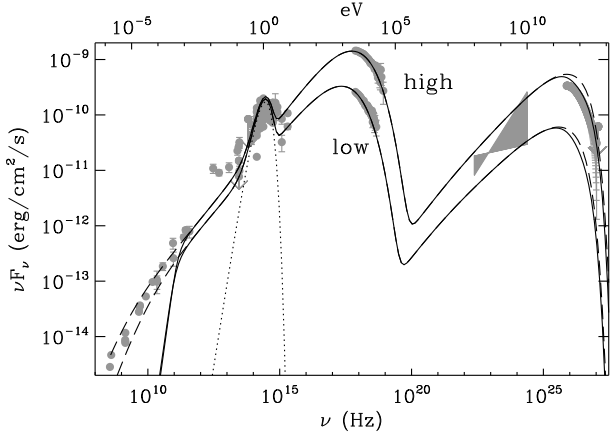}}
\end{center}
\caption{Non-simultaneous spectral energy distribution (SED) of
Mkn~421 with the minimum and maximum X-ray fluxes detected in this
campaign shown. Dashed lines are unabsorbed spectra. Dotted line is
the host galaxy contribution. The `low' and `high' SEDs were fitted by
varying the injected power $P_{\rm inj}$ in the steady-state
homogeneous model, keeping all other parameters identical to those
used for Fig.~\ref{fig:sed2} (see \S\ref{pvar}). The injected particle
energy behaves as $\gamma_b\propto P_{\rm inj}^{1/4}$. These changes
result in wavelength-dependent variations between the two SEDs (see
Fig.~\ref{fig:modrms}). At $\gamma$-ray energies, the SEDs bracket the
long-term average measurement of EGRET. Variations in VHE flux by a
factor $\sim 10$ between low and high states are expected.}
\label{fig:sed1}
\end{figure}

The observed variability decreases very rapidly with wavelength to
only a few percent rms at optical (Fig.~\ref{fig:mwlrms}). X-rays
should therefore provide the best constraints on the necessary
modifications to explain the variability. The spectra corresponding to
the highest and lowest X-ray fluxes measured in the campaign are shown
in Fig.~\ref{fig:sed1}. This figure shows that the simultaneous
detection of Mkn~421 by RXTE and CAT described earlier
(Fig.~\ref{fig:sed2}) occurred while the X-ray flux was quite
close to the maximum observed in this campaign.

A very good representation of the `low' and `high' SEDs is obtained by
varying only the energy of the injected Maxwellian by about 50\% from
$\gamma_b=7.6\times 10^4$ (low) to $11.4\times 10^4$ (high), assuming
that the source has time to reach steady-state (this is relaxed in
\S4.5). The coefficient $K$ (Eq.~\ref{max}) is kept constant to its
value of $3.3\times 10^{-21}$ found from fitting the simultaneous
data. The injected power during the {\em RXTE} observations varies
from 1.1$\times 10^{39}$--5.7$\times 10^{39}$~erg~s$^{-1}$. Hence, the
injection behaves `thermally' with $n_{\rm inj}\propto \gamma_b^3$ or
$P_{\rm inj}\propto \gamma_b^4$. The characteristic $\gamma_b$ is
typically set by the balance of the radiative timescale with the
acceleration timescale $t_{\rm acc}$ but it is unclear how this
relates to $P_{\rm inj}$ without a detailed model for the acceleration
process.

Varying $P_{\rm inj}$ or $n_{\rm inj}$, keeping $\gamma_b$ constant,
alters the normalisation of the synchrotron and IC emission, but does
not reproduce the spectral softening at X-rays with lower
flux. Varying both $\gamma_b$ and $n_{\rm inj}$ (or $P_{\rm inj}$) is
necessary to account for the SEDs. Adjusting both independently, there
being no strong {\em a priori} reason why $n_{\rm inj}\propto
\gamma_b^3$, still leads to this steep dependence.

No other {\em single} parameter change can reproduce the three 
SEDs. It is difficult to fathom why $t_{\rm esc}$, $R$, $B$ or
$\delta$ would change on a timescale $R/c$ in the context of a
one-zone model. In any case, varying these one at a time does not
reproduce the evolution. Combined variations cannot be excluded, but
the simplest solution is that the injected power fluctuates, giving
rise to particles with characteristic energy $\gamma_b\propto P_{\rm
inj}^{1/4}$.

The high and low model SEDs bracket the EGRET long-term average and
have the same spectral slope. The VHE flux varies by an
order-of-magnitude between low and high SEDs. The low VHE flux is
consistent with the lowest measurements reported by Whipple and
HEGRA \citep{1995ApJ...449L..99M,2002A&A...393...89A}.

\begin{figure}[htb!]
\begin{center}
\resizebox{\hsize}{!}{\includegraphics{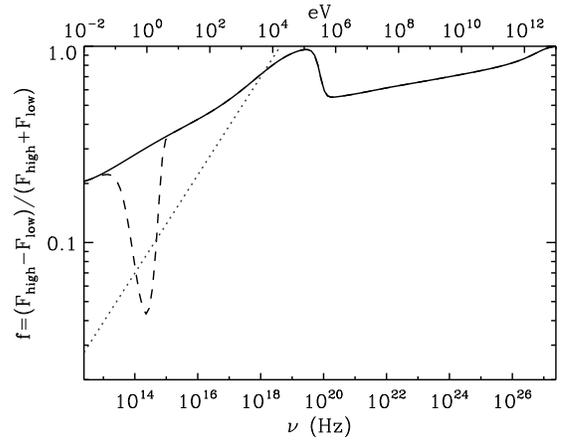}}
\end{center}
\caption{Fractional flux variation between the low and high model SEDs
in Fig.~\ref{fig:sed1}. The dashed line shows the expected dilution in
IR-optical variability if the host galaxy contribution is taken into
account. The dotted line represents the energy dependence of the rms
variability found in Fig.~\ref{fig:mwlrms}. The fractional variation
of the model (which is not an rms) has a shallower slope. The peak
variations do not occur at the synchrotron peak but at the
synchrotron frequency corresponding to the injection $\gamma_b$. IC
becomes dominant above 1 MeV, producing a drop in variability. The IC
variability is less wavelength-dependent since it mirrors changes in
seed photons emitted from UV to X-ray wavelengths. }
\label{fig:modrms}
\end{figure}

Figure \ref{fig:modrms} shows the fractional variation $f=(F_{\rm
high}-F_{\rm low})/(F_{\rm high}+F_{\rm low})$ between the low and
high SEDs. The strongest variability ($f\ga 0.5$) occurs between
1-100~keV and above about 100~MeV (EGRET band). The drop from 0.1 to
1~MeV is due to the increasing dominance of the IC
component. Interestingly, $f$ increases as a power-law from IR to hard
X-rays, just as observed in the rms variability {\em vs} energy plot
(Fig.~\ref{fig:mwlrms}) but with a shallower slope. In particular,
there is no feature associated with the synchrotron peak, which is
around a keV and moves slightly to longer wavelengths with decreasing
flux \citep{2004ApJ...601..759T}.

The variability in the IC component ($\ga$ 1~MeV) is less
wavelength-dependent than the synchrotron component, as it mirrors
primarily the changes in UV to X-ray seed photons. With $f\approx 0.3$
the optical variations are overestimated, although contamination from
the host galaxy could reduce the measured variability from the
expected value. Indeed, such a contribution is included in the plotted
SEDs. It would produce a feature in the rms vs energy plot
(Fig. \ref{fig:modrms}), which seems unlikely since the extrapolation
of the observed X-ray rms is spot on the optical. However, note that
$f$ is not an rms. In principle, one should compute a theoretical
lightcurve by varying $P_{\rm inj}$ according to some prescription and
compare the rms {\em vs} energy theoretical plot (rather than $f$)
with the observed one. Hence, detailed modelling of the rms plot can
constrain how $P_{\rm inj}$ varies. The initial steps are described
below but this is left for future work.

In conclusion, changes between high and low X-ray fluxes can be
modelled as variations in the injected power of the pile-up
distribution.

\subsection{Towards the variability engine\label{engine}}

Until now, particles have been considered to have time to reach
steady-state, thereby neglecting any memory the system might have of
its past condition. More realistically, variations in the injections
will occur on timescales $\la R/c$ in order to reproduce the observed
doubling timescale. Fluctuations in the engine power are mirrored in
the emission of the VHE particles and then propagate down in
wavelength as the particles cool. The system may not have time to
adjust to the new conditions before another variation occurs. The SED
received by the observer will depend on the history of fluctuations in
injection.

\begin{figure}[htb!]
\begin{center}
\resizebox{\hsize}{!}{\includegraphics{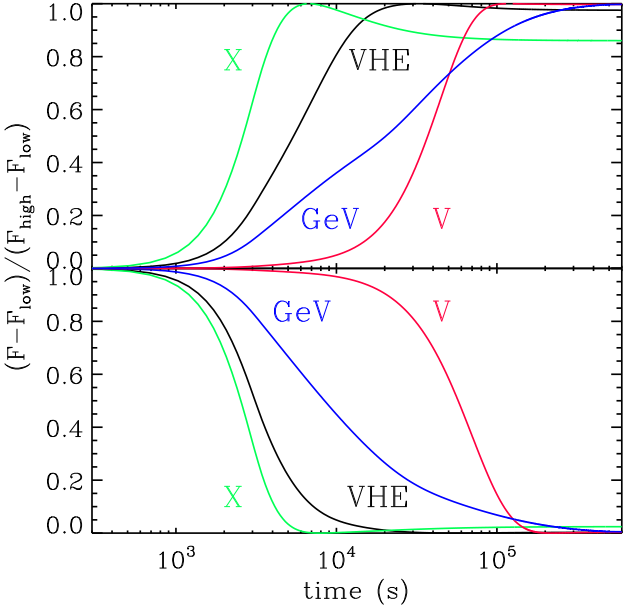}}
\end{center}
\caption{Lightcurve response to a step change in the injection of
particles from low-to-high state (top) and from high-to-low state (bottom). The
injected $\gamma_b$ is abruptly changed at $t=0$ simultaneously
throughout the emission region. Light propagation delays are taken
into account and time is counted in the observer frame. The X-ray
lightcurve is integrated from 2-10~keV, VHE is the integrated flux
above 200~GeV, GeV is the integrated flux between 100~MeV and 10~GeV,
V is the optical $V$ band flux. Note the rapid response of the X-ray
and VHE flux, the $\sim$1 day delay at optical and the gradual
response of the flux in the EGRET band. Correlations between X-ray/VHE
can be expected. However, the long delays and slower rates of change
will hide correlations with the optical or GeV fluxes.}
\label{fig:lightc}
\end{figure}

It is beyond the scope of the present study to consider the resulting
lightcurves for various prescriptions on how $P_{\rm inj}$
varies. Nevertheless, it is possible to illustrate the expected change
in the limiting case of a response to a simultaneous, step-like change
in injection throughout the emission region. Starting with the
steady-state distribution corresponding to the high-state
(resp. low-state) of Fig.~\ref{fig:sed1}, the injection $\gamma_b$ is
abruptly set to its low-state (resp. high-state) value after some
arbitrary time. The resulting lightcurves at various energies are
shown in Fig.~\ref{fig:lightc}. While the change in injection is
simultaneous throughout the emission zone, an observer would see the
impact of the variation first on the closest part of the sphere and
see a change on the opposite side after a time $R/\delta c$ ({\em
e.g.}  \citealt{1999MNRAS.306..551C}).  This light propagation effect
is taken into account when computing the lightcurves, following
\citet{2000ApJ...528..243K}.

As expected, the response is quickest at X-rays and VHE, corresponding
to electrons with the highest energies. This explains the correlations
between these bands. The impact of the change in injection only
propagates down to the optical $V$ band after a delay of $\sim$ a day,
as previously discussed in \citet{1997A&A...320...19M}. This large
delay would explain the lack of correlations between X-rays and
optical bands. The rate of change is a few ks for the X-rays whereas
it takes a few tens of ks for the optical to drop. Sputters in the
variability engine on a timescale $\la 10$~ks will be smoothed out in
the optical compared to X-rays, further reducing the amplitude of the
variations at optical compared to the estimate made in
Fig.~\ref{fig:modrms}. At GeV energies, the variation is also delayed
and very gradual as the IC emission results from an integration over
the seed spectrum. Variations at GeV could be even smoother in time
than at optical wavelengths. Quantitative estimates of these aspects
require a full simulated lightcurve.

\begin{figure}[htb!]
\begin{center}
\resizebox{\hsize}{!}{\includegraphics{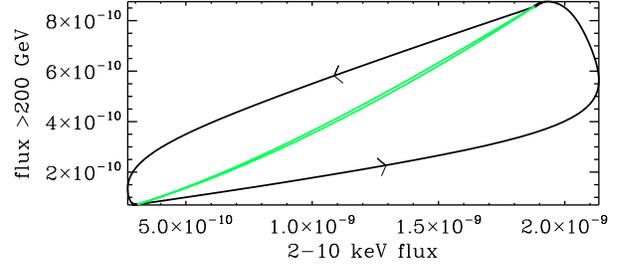}}
\end{center}
\caption{Correlation between X-ray and VHE fluxes plotted in
Fig.~\ref{fig:lightc}. The black curve shows the correlation for step
changes from a low steady-state (bottom left) to a high steady-state
(top right) and back. Arrows indicate the low-to-high and high-to-low
direction in which the loop is traced. The correlation is not
symmetric because of light propagation delays. The grey curve
illustrates the other extreme case in which variations occur on a
timescale $> 10$~ks so that the highest energy particles always
reach steady-state. In this case, the light travel time and cooling
effects have no influence and the change is symmetric. The correlation
is then found to follow $F_{\rm VHE}\propto F_{\rm X}^{1.3}$, as
expected of IC in the Klein-Nishina regime.}
\label{fig:corr}
\end{figure}

The correlation between the X-ray and VHE lightcurves is shown in
Fig.~\ref{fig:corr} for step changes. The finite light travel time and
slightly different cooling combine to produce an asymmetric loop.  At the other extreme, a change in particle
injection could occur at a slow enough pace that the particles are in
steady-state, at least at those energies. Looking at
Fig.~\ref{fig:lightc}, this will typically require variations in the
injection on timescales of many ks. In this case, $P_{\rm inj}$ is
slowly varied between its two extreme values. The flux evolution is
plotted in Fig.~\ref{fig:corr}. Differences between
low-to-high and high-to-low disappear as the SED is in steady-state at
all times. The correlation is well approximated by $F_{\rm VHE}\propto
F_{\rm X}^{1.3}$, as expected for SSC in the Klein-Nishina regime
where $L_{\rm IC}$ is proportional to the number of electrons rather
than its square \citep{1996A&AS..120C.503G}. This is very close to the
long-term trends discussed in \S\ref{sect:var}. Steeper dependences of
the VHE flux with X-ray flux on short timescales, as found in the CAT
data, are not explained to satisfaction
here. The limited dynamic range may bias the
dependences found in datasets with a short time base
\citep{2005A&A...433..479K}. Inversely, these may be able to shed more
light on the workings of the engine than long-term averages. As with
the energy dependence of the rms, computing lightcurves given some
prescription for variations in $P_{\rm inj}$ would be needed to
properly investigate the issue.

\begin{figure}[htb!]
\begin{center}
\resizebox{\hsize}{!}{\includegraphics{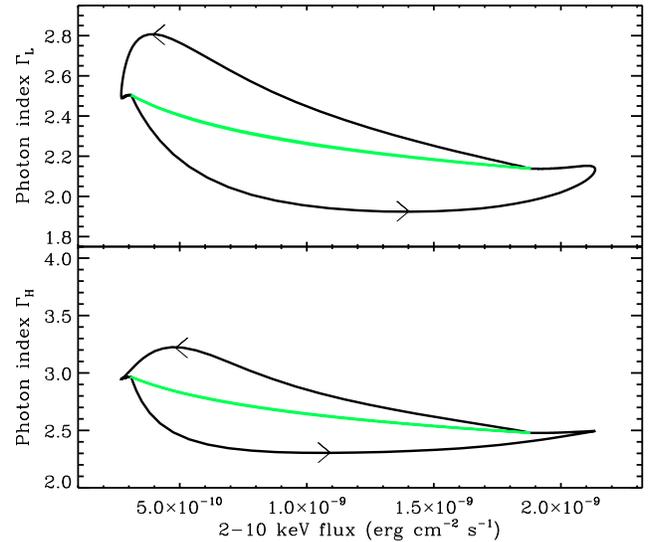}}
\end{center}
\caption{Correlation between X-ray photon indices $\Gamma_{\rm L}$
(top panel) and $\Gamma_{\rm H}$ (bottom panel) with the 2-10~keV
flux. The black solid line corresponds to step changes in the
injection, while the grey solid line is for an injection varying on a
timescale $\gg R/c$ (as in Fig.~\ref{fig:corr}). The
direction of change from low-to-high and back are illustrated by
arrows. The observed indices and fluxes shown in
Fig.~\ref{fig:figs} (on the same scale) fall within the
dark loop.}
\label{fig:index}
\end{figure}

Figure~\ref{fig:index} shows the evolution of the X-ray spectral
indices $\Gamma_{\rm L}$ (2-6.5~keV) and $\Gamma_{\rm H}$ (6.5-10~keV)
in the model for the step-like and slowly varying cases. There is a
flattening above $\sim 10^{-9}$ \flux\ when the variation goes from
low-to-high. There is no such flattening in the opposite
direction. The reasons are not obvious and are related to the
interplay between injection, cooling and light
propagation. Nevertheless, variations in index should occur within
this loop, in agreement with the observations (Fig.~\ref{fig:figs}).

In conclusion, the fits made in \S\ref{pvar} are used to study the
response to variations in the injection from the low to the high-state
values and back. The variations are assumed to be either instantaneous
(step-like) or slow enough for the high-energy electrons to reach
steady-state, thereby illustrating two extreme cases.  Fluctuations in
X-ray and VHE $\gamma$-ray fluxes are found to be correlated with an
index identical to the one derived from long time-base
observations. The expected variations in X-ray photon index {\em vs}
flux also agree well with the observations. The GeV and optical suffer
long delays and respond much more slowly to sudden changes, which
should decrease their variability amplitude and hide correlations with
X-rays or VHE $\gamma$-rays. Hence, the essence of the observational
results garnered in \S3 can be reproduced with a simple change in
injected non-thermal power $P_{\rm inj}$, assuming the distribution is
Maxwellian with a peak energy $\gamma_b\propto P_{\rm
inj}^{1/4}$.

\section{Discussion}
The CAT observations presented here confirm that fast variability on
timescales $\la 1$~hr is a hallmark of VHE emission from
Mkn~421. Nevertheless, the dataset remains limited compared to X-rays,
making the identification of a {\em characteristic} timescale
uncertain. Rapid X-ray flaring, down to sub-hour timescales, appears
to be assocated with high X-ray fluxes -- which is also when Mkn~421
is detected by Cherenkov telescopes \citep{2004ApJ...605..662C}. At
lower fluxes, X-ray variability seems to occur on longer timescales
\citep{2000ApJ...542L.105T}. Observations with long time-bases and
at lower fluxes are needed to characterize better the variability
properties at VHE energies and compare them to the X-ray
properties.

As with other BL Lac objects, the limit from internal absorption of
$\gamma$-rays implies a high relativistic bulk motion in the jet
$\delta> 16$. Even higher values of $\delta\approx 40$ are required to
model the simultaneous CAT, RXTE and optical SED with a single-zone,
homogeneous SSC model. These higher values are needed to ensure that
the emission zone is large enough, hence densities low enough, to
avoid over-producing $\gamma$-ray emission by inverse-Compton
emission. The emission region size $R=2\times 10^{15}$~cm is smaller
and the magnetic field intensity $B=0.2$~G slightly higher than that
found in previous works
\citep{1996ApJ...470L..89T,1997A&A...320...19M,1999ApJ...526L..81M,1999MNRAS.306..551C,2003A&A...410..101K,2003ApJ...597..851K}.

High values for $\delta$ contrast with the kinematics deduced from
radio VLBI observations, which typically imply Doppler factors of a
few at larger scales. VLBA measurements of Mkn 421 realised shortly
after the high VHE state reported here did not find any new components
at the parsec scale after the VHE high state in 2001
\citep{2005ApJ...622..168P}. Solutions to this inconsistency include
strong jet deceleration between the $\gamma$-ray emitting zone and the
parsec scale \citep{2003ApJ...594L..27G}, or a fast-moving spine
surrounded by a slower-moving sheath
\citep{2000A&A...358..104C}. However, high $\delta$ factors are likely
inherent to single-zone SSC models and more complex, inhomogeneous
model can be expected to reduce the discrepancy
\citep{2006ApJ...640..185H}.

The spectral variability of Mkn~421 is investigated under the
assumption of continuous injection of high energy electrons in the
emission region. \citet{1997A&A...320...19M} and
\citet{1999MNRAS.306..551C} had used injection functions $Q\propto
\gamma^{-s} \exp(-\gamma/\gamma_b)$ with $s\approx 1.7$ set in order
to reproduce both the synchrotron X-rays and the power-law at lower
energies. This is a harder value than expected for the power-law
distributions resulting from diffusive shock acceleration.

A Maxwellian injection naturally yields a cooling distribution with an
index $s\approx 1.7$, forfeiting the need to arbitrarily set an
injection index. Pile-ups arise naturally from stochastic acceleration
(second-order Fermi acceleration) on plasma wave turbulence and can be
as effective in producing high energy particles as first-order Fermi
acceleration at shocks
\citep{1984A&A...136..227S,1985A&A...143..431S,1986A&A...162L...1A}.
Quasi-monoenergetic injections may also be expected from reconnection
in Poynting-flux dominated jets ({\em e.g.}
\citealt{2005ApJ...625...72S} and references therein).

The substantial X-ray spectral variations are well reproduced by a
factor $\approx$ 5 change in injected power. The characteristic energy
of the particles $\gamma_b$ varies as $P_{\rm inj}^{1/4}$. A
thermal-like correspondence between total injected power and
characteristic energy of the VHE particles is surprising: even if the
acceleration process results in a pile-up distribution, why should the
injection rate vary as $\gamma_b^3$ ?  Whether this is a coincidence
or something which to be addressed by acceleration theory remains to
be decided.

The injected particles have a characteristic $\gamma_b\sim 10^5$, which 
corresponds to the minimum energy for which electrons cool in a time $\la 
R/c$, and which also correponds to the maximum energy that can be reached 
in the acceleration
process before cooling dominates. This suggests that $t_{\rm acc}\sim 
t_{\rm cool}\sim R/c$. Symmetric rise and decay lightcurves are expected 
in this case at X-ray energies \citep{1999MNRAS.306..551C}, as observed in 
Mkn~421 \citep{2004ApJ...605..662C}.

The higher $\gamma_b$ hints at a 
decreasing $t_{\rm acc}$, hence a more efficient particle acceleration 
mechanism in the high state. This also explains the saturation of the 
X-ray spectral indices above a certain flux. The index varies less when the synchrotron peak reaches the spectral band. Such a saturation has also 
been observed in Mkn~501. A detailed comparison between these and other 
sources could provide simpler diagnostics of acceleration at the highest 
energies than hysteresis loops ({\em e.g.} 
Fig.~\ref{fig:corr}). Both clockwise and counter-clockwise loops have been 
seen in Mkn~421 
\citep{1996ApJ...470L..89T,2000ApJ...541..166F,2004ApJ...605..662C}. The 
choice of spectral bands and the base-line emission may significantly 
affect the resulting loops and make them ambiguous to interpret as 
isolated cooling/heating events, whereas an overall flux - index plot for 
the whole dataset may provide a broader, simpler picture to address.

Likewise, the power-law dependence of the fractional variability with 
energy from optical to X-ray energies may offer new insights into the 
workings of the engine. It does not imply correlations between the 
lightcurves at all these energies ({\em i.e.} the phase information can be 
lost). Indeed, the relationship between rms variability and energy seems unlinked to the flux level.  In 
the present model, the engine must somehow be able to wag the high energy 
tail of the particle distribution in the right
way so that the variation propagates down in energy with a 
tightly-constrained decreasing amplitude. Theoretical lightcurves, obtained by
varying $P_{\rm inj}$, should be computed and their rms compared to
the observations. Interestingly, the slow response of the GeV flux to
step changes in injection suggests strong variations at X-ray and TeV
energies which may not be accompanied by similar variations at GeV energies. Characterizing the amplitude and correlation with other bands by detailed modelling will be useful for {\em GLAST}.

\section{Conclusion}
Simultaneous observations carried out with CAT and {\em RXTE} confirm the 
correlation between X-ray and VHE emission and the fast sub-hour 
variability at VHE energies. A saturation of the X-ray spectral indices 
with X-ray flux and a power-law relationship between fractional 
variability and energy (from optical to X-rays) are found. These results 
may extend to other TeV blazars, and thereby provide strong, universal 
constraints on models of high energy emission from relativistic jets.

The SED and variability are well reproduced by a single-zone,
homogeneous model in which a Maxwellian distribution of electrons with
$\gamma_b\sim 10^5$ are continuously injected. Such an injection would
favour second-order Fermi acceleration or neutral point acceleration
over the more widely studied shock acceleration. Cooling by
synchrotron self-Compton emission in the Klein-Nishina regime
naturally leads to a steady-state distribution of electrons with the
correct slope to explain the emission down to radio frequencies. The
X-ray spectral saturation and power-law rms spectrum are reproduced by
a factor 5 change in the power $P_{\rm inj}\propto \gamma_b^{1/4}$ of
the injected Maxwellian. The acceleration process is probably
responsible for this dependence, while variations in $P_{\rm inj}$
depend on the workings of the internal engine. Detailed lightcurves,
particularly at $\gamma$-ray energies where the information is sparse,
can therefore provide important insights into the mechanisms at work
in blazars.

\begin{acknowledgements}

We thank Kari Nilsson and the Tuorla Observatory astronomers who have
worked on the Mkn~421 observations for the optical lightcurve.  

We acknowledge the CAT collaboration for the permission to use
archival results, the RXTE Science Operations Center staff
providing the observations and the RXTE Guest Observer Facility for
providing support in analyzing them.

This research has made use of the NASA/IPAC Extragalactic Database
(NED) which is operated by the Jet Propulsion Laboratory, California
Institute of Technology, under contract with the National
Aeronautics and Space Administration.
\end{acknowledgements}

\bibliography{lsrefs,myref}

\end{document}